\documentclass[12pt,preprint]{article}
\usepackage{amssymb}
\usepackage{amsmath}
\usepackage{graphics}
\usepackage{epsfig}

\newcommand{\eqb}{\begin{equation}}
\newcommand{\eqe}{\end{equation}}
\newcommand{\dmb}{\begin{displaymath}}
\newcommand{\dme}{\end{displaymath}}

\newcommand{\eab}{\begin{eqnarray}}
\newcommand{\eae}{\end{eqnarray}}

\newcommand{\be}{\begin{equation}}
\newcommand{\ee}{\end{equation}}

\input{tcilatex}

\begin{document}

\begin{titlepage}
\begin{flushright} 
\end{flushright}
\vspace{0.6cm}

\begin{center}
\Large{A Planck-scale axion and SU(2) 
Yang-Mills dynamics: Present acceleration 
and the fate of the photon}
\vspace{1.5cm}

\large{Francesco Giacosa and Ralf Hofmann}

\end{center}
\vspace{1.5cm} 

\begin{center}
{\em Institut f\"ur Theoretische Physik\\ 
Universit\"at Frankfurt\\ 
Johann Wolfgang Goethe - Universit\"at\\ 
Max von Laue--Str. 1\\ 
60438 Frankfurt, Germany}
\end{center}
\vspace{1.5cm}

\begin{abstract}

From the time of CMB decoupling onwards we 
investigate cosmological evolution subject to a strongly 
interacting SU(2) gauge theory of Yang-Mills scale 
$\Lambda\sim 10^{-4}$\,eV (masquerading as the $U(1)_{Y}$ factor of the SM at present).
The viability of this postulate is discussed in 
view of cosmological and (astro)particle physics bounds. The gauge theory 
is coupled to a spatially homogeneous and ultra-light (Planck-scale) axion field. 
As first pointed out by Frieman et al., such an axion is a viable candidate for 
quintessence, i.e. dynamical dark energy, being 
associated with today's cosmological acceleration. A prediction
of an upper limit $\Delta t_{m_\gamma=0}$ for the duration
of the epoch stretching from the present 
to the point where the photon starts to be Meissner 
massive is obtained: $\Delta t_{m_\gamma=0}\sim 2.2$ billion years.
        
\end{abstract} 

\end{titlepage}

\bigskip

\section{Introduction}

The possibility to interpret dark energy in terms of an ultra-light
pseudo-Nambu-Goldstone boson field is at the center of an exciting debate
stretching over the last decade, see e.g. \cite%
{frieman1995,kim2002,wilczek2004,hall2005,kaloper2005,Barbieri:2005gj}. The
idea is that an axion field $\phi $, which is generated by Planckian
physics, develops a small mass due to topological defects of a Yang-Mills
theory. If the associated Yang-Mills scale is far below the Planck mass $%
M_{P}$ then $\phi $'s slow-roll dynamics at late time can mimic a small
cosmological constant being in agreement with the present observations.
Having the coherent field $\phi $ decay by increasingly efficient
self-interactions at late time, the associated very light pseudoscalar
bosons interact with ordinary matter only very weakly and thus escape their
detection in collider experiments. Because of a dynamically broken, global
U(1)$_{A}$ symmetry associated with the very existence of $\phi$ the
corresponding potential $V(\phi )$ is radiatively protected. Notice that $%
V(\phi )$ is generated by an explicit, anomaly-mediated breaking of U(1)$%
_{A} $.

Up to small corrections, arising from multi-instantons effects, $V(\phi)$
has the following form \cite{PecceiQuinn1977} 
\begin{equation}
V(\phi )=\mu ^{4}\left[ 1-\cos \left(\frac{\phi}{F}\right) \right]\,.
\label{pot}
\end{equation}%
Two mass scales enter in eq.\,(\ref{pot}): the dynamical symmetry breaking
scale $F$ (axion decay constant) and a scale $\mu$ associated with the
explicit symmetry breaking. The scale $\mu$ roughly determines at what
momentum scale the gauge theory providing the topological defects becomes
strongly interacting. Recall that the potential (\ref{pot}) is an effective
one, arising from a quantum anomaly of the U(1)$_A$ symmetry which is
defined on integrated-out fermion fields. The anomaly becomes operative
through topological defects of a Yang-Mills theory and is expressed by an
additional, CP violating contribution 
\begin{equation}
\mathcal{L}_{\phi-\mbox\tiny{SU(2)}}=\frac{\phi }{32\pi ^{2}F}F_{\mu \nu
}^{a}\widetilde{F}^{a,\mu \nu }\,.  \label{lag}
\end{equation}%
Upon integrating over topological sectors, one concludes that the parameter $%
\mu$ in (\ref{pot}) is comparable to the Yang-Mills scale $\Lambda$ \cite%
{PecceiQuinn1977,Kim:2003se}.

The mass $m_{\phi}$ of the field $\phi$, as derived from (\ref{pot}) for the
range $\left\vert \phi \right\vert \lesssim \frac{\pi}{2} F$, reads $m_{\phi
}\simeq \mu ^{2}/F$. Assuming $\phi\sim F\sim 10^{18}\,$GeV, the needed
value for $\mu$ to generate the present density of dark energy in the
universe (inferred from fits to SNe Ia luminosity distance--redshift data
for $z<1.7$ \cite{Riess1998,Perlmutter1998,Schmidt1998}) is $\mu\sim 10^{-3}$%
\,eV \cite{frieman1995}. By the closeness of $\mu$'s value to the MSW
neutrino mass a possible connection with neutrino physics was suggested in 
\cite{frieman1995} (see also \cite{Barbieri:2005gj}).

In the present work we wish to propose a different axion-based scenario
relating the (presently stabilized) temperature of the CMB, $T_{{\tiny %
\mbox{CMB}}}=2.35\times 10^{-4}$\thinspace eV, with the present scale of
dark energy $\sim 10^{-3}\,$eV. Namely, we postulate that the gauge factor $%
U(1)_{Y}$ of the standard model of particle physics (SM) is only an
effective manifestation of a larger gauge group. According to \cite%
{Hofmann2005} one is lead to consider SU(2)$\supset U(1)_{Y}$\footnote{%
For the present discussion we disregard the fact that in the SM the unbroken
generator, corresponding to U(1)$_{\tiny\mbox{em}}$, is a linear combination,
parametrized by the Weinberg angle, of the diagonal SU(2)$_{W}$'s generator
and the U(1)$_{Y}$ generator in unitary gauge since we are not concerned
with the interactions of the photon with electrically charged leptons and/or
hadrons which would make this mixing operative. Our investigation of
cosmology below sets in at the point where the CMB decoupling takes place.
The issue is, however, re-addressed in Sec.\thinspace \ref{viab}.}
(henceforth referred to as SU(2)$_{{\tiny \mbox{CMB}}}$) as a viable
candidate for such an enlargement of the SM's gauge symmetry. In spite of
the fact that such a postulate is rather unconventional we nevertheless feel
that a fruitful approach to the dark-energy problem needs novel Ans\"{a}tze.
In a slightly different context a QCD-like force of scale $\sim 10^{-3}\,$eV
was also discussed in \cite{hall2005}.

We intend to explore some consequences of the postulate SU(2)$_{{\tiny %
\mbox{CMB}}}\overset{{\tiny \mbox{today}}}{=}$U(1)$_{Y}$ in connection with
axion physics. The observation of a massless and unscreened photon strongly
constrains the region in the phase diagram of the SU(2) Yang-Mills theory
corresponding to the present state of the Universe \cite%
{Hofmann2005,HerbstHofmannRohrer2004}. As a consequence, the Yang-Mills
scale $\Lambda _{{\tiny \mbox{CMB}}}$ is determined to be comparable to $T_{%
{\tiny \mbox{CMB}}}$: $\Lambda _{{\tiny \mbox{CMB}}}\sim 10^{-4}$ eV.

The fact that $T_{{\tiny \mbox{CMB}}}$ is comparable to the Yang-Mills scale
of a theory with gauge group SU(2)$_{{\tiny \mbox{CMB}}}$ (containing the
U(1)$_Y$ factor of the SM as a subgroup) which in connection with a
Planck-scale axion field explains the present density of dark energy, does
by itself not constitute a proof for the existence of SU(2)$_{{\tiny %
\mbox{CMB}}}$ in Nature. For this setup to be convincing it ought to make
independent and experimentally verified pre- and postdictions such as a
dynamical account of the large-angle features of CMB maps induced by the
nonabelian fluctuations of SU(2)$_{{\tiny \mbox{CMB}}}$. In this sense the
present paper is the very first stage in a long-term program exploring the
implications of SU(2)$_{{\tiny \mbox{CMB}}}\overset{{\tiny \mbox{today}}}{=}$%
U(1)$_{Y}$, see also \cite{Hofmann22005}. The authors are well aware of the
fact that this program may lead to the falsification of the postulate SU(2)$%
_{{\tiny \mbox{CMB}}}\overset{{\tiny \mbox{today}}}{=}$U(1)$_{Y}$. Mounting
evidence for its correctness is, however, provided by two-loop calculations
of thermodynamical quantities in the deconfining phase of SU(2) Yang-Mills
theory, see \cite{HerbstHofmannRohrer2004,SHG2006}, making a further
pursuit of this program worthwhile.

For the reader's convenience, let us put the results of ref.\thinspace \cite%
{Hofmann2005} into perspective with other approaches to Yang-Mills
thermodynamics. First of all, it is important to note that ref.\thinspace 
\cite{Hofmann2005} considers the case of pure thermodynamics only, that is,
the absence of external (static or dynamic) sources which would upset the
spatial homogeneity of the system. As a consequence, the approach in \cite%
{Hofmann2005} has nothing to say about the spatial string tension in the
deconfining phase. The spatial string tension, introducing a distance scale $%
R$ into the system, can, however, be easily extracted in lattice simulations
of the spatial Wilson loop \cite{KorthalsAltes,HoelbingRebbiRubakov2001}. It
is conceivable that static sources can be treated adiabatically based on the
approach of \cite{Hofmann2005} by assuming a position dependence of
temperature, but this is the subject of future research. In the absence of
sources, a situation that is of relevance for physics on cosmological length
scales, refs.\thinspace \cite{Hofmann2005,Hofmann2006} give a detailed and
reliable account of SU(2) and SU(3) Yang-Mills thermodynamics. Lattice
simulations for SU(3), using the differential method \cite%
{Brown1988,Deng1988}, which is adopted to finite lattice sizes, yield
quantitative agreement with the result for the entropy density (an infrared
safe quantity) obtained in ref.\thinspace \cite{Hofmann2005}. The results
for the pressure and the energy density (infrared sensitive) agree
qualitatively with those obtained in \cite{Brown1988,Deng1988}. That is, in
contrast to the integral method, which assumes the infinite-volume limit on
a finite-size lattice, the pressure is negative shortly above the critical
temperature in the deconfining phase, and there is a power-like fast
approach to the Stefan-Boltzmann limit for $T\rightarrow \infty $. Both
phenomena are observed in the approach of \cite{Hofmann2005}, but
numerically the results differ in the vicinity of the phase boundary%
\footnote{%
We believe that this is an artefact of the finite lattice volume which close
to the phase boundary affects the long-range correlations present in the
ground state. Hence we dismiss the widely used argument that the imprecise
knowledge of the lattice $\beta $-function would be the cause of the \textsl{%
apparent} problems with the differential method.}. By virtue of the trace
anomaly the gluon condensate, up to a factor weakly depending on temperature
($\beta $-function over fundamental coupling $g$), coincides with the trace
of the energy-momentum tensor $\rho -3P$. Neglecting the masses and
interactions of the excitations, which is an excellent approximation at high
temperatures as far as the excitation's equation of state is concerned, we
have $\rho -3P\propto T$. In \cite{Langfeld} the temperature dependence of
the SU(2) gluon condensate was investigated on the lattice, and, indeed a
linear rise of the gluon condensate with temperature was observed. Notice
the conceptual and technical differences of \cite{Hofmann2005} to the
hard-thermal-loop (HTL) approach \cite{HTL}. The latter derives a nonlocal
theory for interacting soft and ultrasoft modes. While the HTL approach, in
a highly impressive way technically, integrates perturbative ultraviolet
fluctuations into effective vertices it can not shed light on the
stabilization of the infrared physics associated with nonperturbative
fluctuations residing in the magnetic sector of the theory. The derivation
of the phase $\phi /|\phi |$ in \cite{Hofmann2005,HerbstHofmann2004},
however, invokes these nonperturbative, magnetic correlations. An impressive
machinery has been developed within the renormalization-group flow approach
to Yang-Mills thermodynamics both in the imaginary \cite{Litim} and the
real-time approach \cite{Pietroni} the latter making important observations
concerning the (nonperturbative) temperature dependence of the thermal gluon
mass. Interesting results for the nonperturbative temperature dependence of
the fundamental gauge coupling in Quantum Chromodynamics were obtained in 
\cite{Gies}. However, to the best of the authors knowledge, no decisive
nonperturbative calculation (fully considering the magnetic sector) of the
thermodynamical pressure or related quantities has yet been performed within
this approach.

The paper is organized as follows: First, we briefly recall some basic
nonperturbative results obtained in \cite{Hofmann2005} for SU(2) Yang-Mills
thermodynamics. Then we discuss the viability of our scenario when
confronting it with particle-physics experiments and cosmological
observations. Subsequently, we consider for a spatially flat Universe the
evolution of the cosmological scale factor $a=a(t)$ and of the axion-field $%
\phi =\phi (t)$ from the time of decoupling $t_{dec}$ (corresponding to $%
z_{dec}=1089$) up to the present. We then investigate the future evolution
of the Universe up to the point when the transition between the deconfining
and the preconfining phase of SU(2)$_{{\tiny \mbox{CMB}}}$ will take place
and the photon will acquire a Meissner mass. Finally, we present our
conclusions and an outlook on future research.

\section{SU(2) Yang-Mills thermodynamics\label{dp}}

In \cite{Hofmann2005} a nonperturbative approach for SU(2)/SU(3) Yang-Mills
thermodynamics is developed. For the sake of brevity we recall only some of
the results relevant for the present study. Analytical expressions are
reported in the Appendix.

\noindent\textit{Deconfining (electric) phase:}

\begin{figure}[tbp]
\begin{center}
\leavevmode
\leavevmode
\vspace{4.3cm} \includegraphics{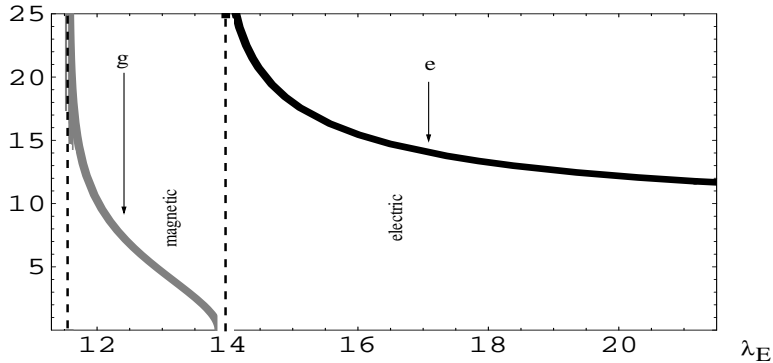}
\end{center}
\caption{{Temperature evolution of the effective (electric and magnetic) couplings.}}
\label{Fig1B}
\end{figure}

At high $T$ it is shown in \cite{Hofmann2005, HerbstHofmann2004} that an
inert adjoint Higgs field $\varphi $ emerges upon spatial coarse-graining
over topological defects. The modulus of the Higgs field is
(nonperturbatively) temperature dependent with $\left\vert \varphi
_{E}\right\vert =\sqrt{\Lambda _{E}^{3}/2\pi T}$ (here $\Lambda _{E}$
denotes the Yang-Mills scale as defined in the deconfining phase \cite%
{Hofmann2005}); furthermore, the field $\varphi _{E}$ induces a dynamical
gauge symmetry breaking SU(2)$\rightarrow $ U(1): two of the three gauge
bosons become massive (denoted by $V^{\pm }$), while the third one remains
massless (denoted by $\gamma $). Massive excitations $V^{\pm }$ are very
weakly interacting thermal quasi-particles, their mass depends on
temperature as%
\begin{equation}
m_{V^{\pm }}=2e(T)\left\vert \varphi _{E}\right\vert   \label{mvpvm}
\end{equation}%
where $e(T)$ is an \textsl{effective} temperature-dependent gauge coupling as 
plotted in Fig.\,\ref{Fig1B}.
The gauge coupling $e(T)$ diverges logarithmically at $T_{c,E}$
and has a plateau value of $e\simeq 8.89$ for $T\gg T_{c,E}$. For $T\searrow
T_{c,E}$ the $V^{\pm }$ bosons acquire an infinite mass, $%
m_{V^{+}}=m_{V^{-}}\propto -\log (T_{E}-T_{c,E})$. Thus they do no longer
(weakly) screen the propagation of the massless excitation $\gamma $.

\begin{figure}[tbp]
\begin{center}
\leavevmode
\leavevmode
\vspace{4.3cm} \includegraphics{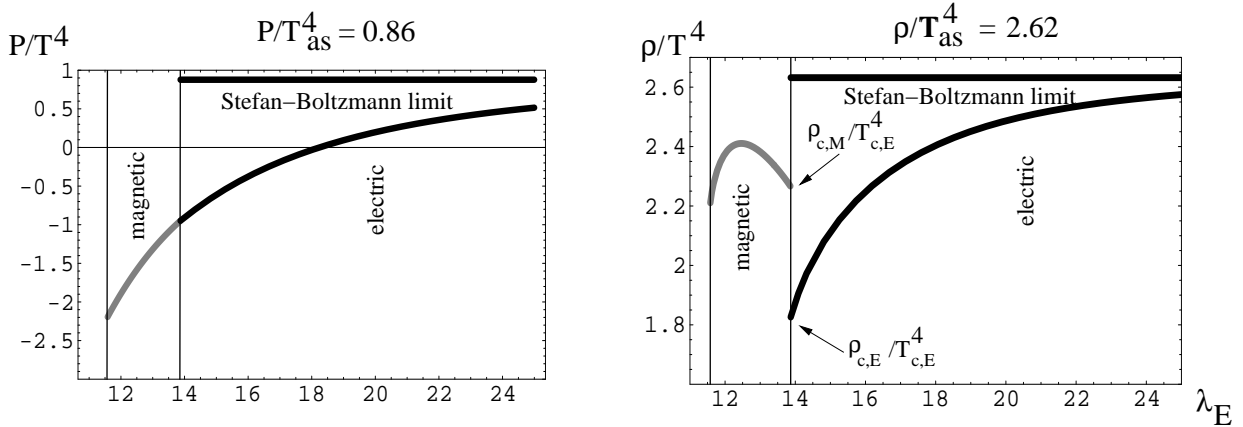}
\end{center}
\caption{{Ratios of Pressure (left panel) and energy density (right panel)
to $T^{4}$ of an SU(2)Yang-Mills theory within the deconfining (electric) and
preconfining (mangetic) phases. The vertical lines denote phase boundaries. }}
\label{Fig1}
\end{figure}

Plots for the energy-density and for the pressure as functions of the
dimensionless temperature $\lambda _{E}=\frac{2\pi T}{\Lambda _{E}}$ are
shown in Fig.\thinspace\ref{Fig1}. 

In Fig.\thinspace \ref{Fig1} a jump in the energy density at the critical
value $\lambda _{c,E}=13.87$ is seen. This signals a transition between the
deconfining (also called electric, $T\geq T_{c,E}$) and the preconfining
(also called magnetic, $T_{c,M}\leq T<T_{c,E}$) phases. Notice that for high 
$T$ the Stefan-Boltzmann limit is approached in a power-like fashion.

\noindent\textit{Preconfining (magnetic) phase:}

In the preconfining phase the dual gauge symmetry $U(1)$ is dynamically
broken by a magnetic monopole condensate, which, after spatial
coarse-graining, is described by an inert complex scalar field $\varphi _{M}$
with $\left\vert \varphi _{M}\right\vert =\sqrt{\Lambda _{M}^{3}/2\pi T}.$
The dual gauge excitation $\gamma $ now acquires a Meissner mass 
\begin{equation}
m_{\gamma }=g(T)\left\vert \varphi _{M}\right\vert ,
\end{equation}%
where $g(T)$ is the effective coupling in the magnetic phase as shown 
in Fig.\,\ref{Fig1B}. The latter
vanishes for $T\nearrow T_{c,E}$ and it diverges logarithmically for For $%
T\searrow T_{c,M},$ where $T_{c,M}\sim 0.83\,T_{c,E}$ in the SU(2) case.

For $T\nearrow T_{c,E}$ a jump in the number of polarizations ($3\rightarrow
2$) takes place thus explaining the discontinuity $\rho _{c,M}\rightarrow
\rho _{c,E}$ in the energy density, see Fig.\thinspace \ref{Fig1}. Due to
the dominance of the ground state the pressure is negative for $T\sim T_{c,E}
$ (for a microscopic explanation of this macroscopically stabilized
situation see \cite{Hofmann2005}).

For $T\searrow T_{c,M}$ the excitation $\gamma $ becomes infinitely massive,
thus signalling a second phase transition which is of the Hagedorn type.

\noindent\textit{Confining (center) phase:}

For temperatures below $T_{c,M}$ the system is in its confining phase:
fundamental test-charges and gauge modes are confined and decoupled,
respectively. The excitations are (single and selfintersecting) fermionic
center-vortex loops \cite{Hofmann2005}.

When confronting these results with the postulate SU(2)$_{{\tiny \mbox{CMB}}}%
\overset{{\tiny \mbox{today}}}{=}$U(1)$_{Y}$ the reader may be puzzled by
the existence of two massive excitations in addition to the photon for $T>T_{%
{\tiny \mbox{CMB}}}$; however, as shown in \cite{Hofmann2005} and explained
in more detail in Sec.\thinspace \ref{viab}, the interaction between $\gamma 
$ and $V^{\pm }$ is tiny because the off-shellness of admissible quantum
fluctuations is strongly constrained by the applied spatial coarse-graining.
At $T=T_{c,E}=T_{{\tiny \mbox{CMB}}}$, where the $V^{\pm }$ disappear from
the thermal spectrum, $\gamma $ is exactly noninteracting. As an
experimental fact today's on-shell photons are massless on the scale of $T_{%
{\tiny \mbox{CMB}}}=2.35\times 10^{-4}$\thinspace eV ($m_{\gamma }<10^{-13}\,
$eV \cite{Williams1971}), and they are unscreened. As a consequence, the
postulated SU(2)$_{{\tiny \mbox{CMB}}}$ dynamics necessarily is
characterized by a temperature $T_{c,E}\sim T_{{\tiny \mbox{CMB}}}$ today:
SU(2)$_{{\tiny \mbox{CMB}}}\overset{{\tiny \mbox{today}}}{=}U(1)_{Y}$. This
entails $\Lambda _{{\tiny \mbox{CMB}}}\sim 1.065\times 10^{-4}\,$eV and a
present SU(2)$_{{\tiny \mbox{CMB}}}$ ground-state pressure $P^{g.s.}\sim
-(2.44\times 10^{-4}\,\mbox{eV})^{4}$. Before discussing the phenomenological
and cosmological viability of SU(2)$_{{\tiny \mbox{CMB}}}\overset{{\tiny %
\mbox{today}}}{=}$U(1)$_{Y}$, two comments are in order:

(i) For $T\searrow T_{c,E}$, the SU(2)$_{{\tiny \mbox{CMB}}}$ system will
not immediately jump to the preconfining phase because of the discontinuity
in the energy density, see Fig.\thinspace \ref{Fig1} and the corresponding
evaluation in the Appendix. Rather, it remains in a supercooled state until
a temperature $T_{\ast }\in (T_{c,M},T_{c,E})$ is reached where a
restructuring of the ground state (interacting calorons $\rightarrow $
interacting, massless monopoles \cite{Hofmann2005}) does not cost any
energy. For a detailed discussion of this situation see Sec.\thinspace \ref%
{masspho}.

(ii) If SU(2)$_{{\tiny \mbox{CMB}}}\overset{{\tiny \mbox{today}}}{=}U(1)_{Y}$
strictly holds then it was not always so: for $T>T_{{\tiny \mbox{CMB}}}$
photons did interact with the massive partners $V^{\pm }$. Nonabelian
effects peak at $T\sim 3\,T_{c,E}$ and are visible on the level of $10^{-3}$
in the relative deviation from the ideal photon-gas pressure, see \cite{HerbstHofmannRohrer2004,SHG2006}. This represents a crucial test for our basic
postulate. We suspect that the effect generates a dynamical contribution to
the dipole of the CMB temperature map in addition to a component induced by
the relativistic Doppler effect \cite{PW1968}. A more quantitative analysis
of this assertion is beyond the scope of the present paper but is planned as
a next step.

\section{But is it viable?\label{viab}}

To see that the suggestion SU(2)$_{{\tiny \mbox{CMB}}}\overset{{\tiny %
\mbox{today}}}{=}$U(1)$_{Y}$ is viable when confronted with observational
facts in cosmology (nucleosynthesis) and (astro)particle-physics bounds on
neutral and charged current interactions we need to consider the following
points: \textsl{(a)} Big Bang Nucleosynthesis (BBN) bounds on the number of
relativistic degrees of freedom at $T\sim 1$ MeV in the SM. To resolve an
apparent contradiction with SU(2)$_{{\tiny \mbox{CMB}}}\overset{{\tiny %
\mbox{today}}}{=}$U(1)$_{Y}$ we discuss the underlying gauge dynamics of the
weak sector which now is based on \textsl{pure} SU(2) Yang-Mills dynamics.
The latter entails results such as a higgs-particle free and stepwise
mechanism for electroweak symmetry breaking and the dynamical emergence of
the lepton families. \textsl{(b)} Constraints on the $\gamma $-$V^{\pm }$
interaction for $T>T_{{\tiny \mbox{CMB}}}$. \textsl{(c)} Charged currents.
This is of relevance for the discussion of supernova cooling. \textsl{(d)} $%
\gamma$ interaction with charged leptons and neutral currents (particle-wave
duality of the photon).

\noindent\textit{(a) Numbers of relativistic degrees of freedom.}

Relying on the observed primordial $^{4}$He and D abundances and on a baryon
to photon number density ratio $\eta $ ranging as $4.9\times 10^{10}\leq
\eta \leq 7.1\times 10^{10}$ SM based nucleosynthesis predicts that the
number of relativistic degrees of freedom $g_{\ast }$ at the freeze-out
temperature $T_{{\tiny \mbox{fr}}}\sim 1\,$MeV is given as 
\begin{equation}
g_{\ast }=5.5+\frac{7}{4}N_{\nu }  \label{reldof}
\end{equation}%
with $1.8\leq N_{\nu }\leq 4.5$ \cite{Cyburt2005}. This prediction relies on
the following argument: the neutron to proton fraction $n/p$ at freeze-out
is given as $n/p=\exp [-Q/T_{{\tiny \mbox{fr}}}]\sim 1/6$ where $Q=1.293\,$%
\thinspace MeV is the neutron-proton mass difference and one has 
\begin{equation}
T_{{\tiny \mbox{fr}}}\sim \left( \frac{g_{\ast }G_{N}}{G_{F}^{4}}\right)
^{1/6}\,.  \label{Tfreeze}
\end{equation}%
In eq.\thinspace (\ref{Tfreeze}) $G_{N}$ denotes the Newton constant, and 
\begin{equation}  \label{GF}
G_{F}=\pi \frac{\alpha _{w}}{\sqrt{2}\,m_{W}^{2}} \sim 1.17\times 10^{-5}\,%
\mbox{GeV}^{-2}
\end{equation}
is the Fermi coupling at zero temperature. To use the zero-temperature value
of $G_{F}$ at $T_{{\tiny \mbox{fr}}}=1$\,MeV, as it is done in eq.\,(\ref%
{Tfreeze}), is justified by the large `electroweak scale' $v=247$\thinspace
GeV: the vacuum expectation of the fundamentally charged Higgs-field in the
SM. SU(2)$_{{\tiny \mbox{CMB}}}\overset{{\tiny \mbox{today}}}{=}$U(1)$_{Y}$ 
tells us that there are effectively six
relativistic degrees of freedom at $T_{{\tiny \mbox{fr}}}=1\,$MeV in
addition to the situation described by the SM: a result which clearly
exceeds the above cited upper bound for $N_{\nu }$. But does this falsify
our postulate SU(2)$_{{\tiny \mbox{CMB}}}\overset{{\tiny \mbox{today}}}{=}$%
U(1)$_{Y}$ or is there new physics associated with the thermalization of the
weak sector of the SM? In what follows we will argue that the approach to
Yang-Mills theory sketched in Sec.\,\ref{dp} should also be applied to the
electroweak group SU(2)$_{W}$ (which we refer to as SU(2)$_{e}$ where $e$
stands for `electron', see below). In \cite{Hofmann2005} we have discussed
why and how the assignment SU(2)$_W=$SU(2)$_e$ (the associated Yang-Mills
scale is $\Lambda_e\sim m_e\sim 0.5$\,MeV) works to generate a triplet of
intermediary massive vector bosons: $W^\pm$ decouples at the
deconfining-preconfining phase boundary $T=T_{c,E}$ (second-order like
transition) while $Z_0$ decouples at the boundary between preconfining and
confining phase $T=T_{c,M}$ (Hagedorn transition). Thus the weak symmetry is
broken in a stepwise fashion (notice that $\Lambda_e\sim m_e\sim T_{c,M}\sim
0.835\,T_{c,E}$). Moreover, the first lepton family $(e,\nu_e)$ emerges in
the confining phase of SU(2)$_e$ (single and selfintersecting center-vortex
loop; all higher-charge states are unstable). The effective $V-A$ structure
of the weak currents likely emerges as a consequence of the departure from
(local) thermal equilibirum close to the Hagedorn transition causing the CP
violating Planck-scale axion to fluctuate. The mass of the selfintersecting
center-vortex loop $m_e$ (mass of the electron) is roughly equal to the
scale $\Lambda_e$. Furthermore, the single center-vortex loop emerges as a
Majorana particle \cite{Hofmann2005} in agreement with experiment \cite%
{Klapdor2004}. As another consequence of SU(2)$_W=$SU(2)$_e$, the hierarchy $%
g^{-1}_{{\tiny \mbox{dec}}}\equiv g^{-1}(T_{c,M})\sim\frac{m_{e}}{m_{Z_{0}}}%
\sim 10^{-5}\sim \frac{m_{\nu _{e}}}{m_{e}}$ is \textsl{not} explained by
the large value $v$ of the Higgs expectation, a moderate value of the gauge
coupling, and a $B-L$ forbidden neutrino mass but by the logarithmic pole $%
g\sim-\log(T-T_{c,M})$ of the magnetic coupling \cite{Hofmann2005}. Thus the
high-energy scale associated with the Higgs sector of the SM turns out to
emerge in terms of \textsl{pure} but nonperturbative Yang-Mills physics.

Now at $T_{{\tiny \mbox{fr}}}=1\,$MeV SU(2)$_{e}$ is in its deconfining
phase. As a consequence, $Z_{0}$ is massless at $T_{{\tiny \mbox{fr}}}$ and
the mass of $W^{\pm }$ is substantially reduced compared to its decoupling
value \cite{Hofmann2005}. The latter, in turn, is responsible for the
zero-temperature value of $G_F$, see eq.\,(\ref{GF}). This, however, implies
that the $G_F$ in eq.\thinspace (\ref{Tfreeze}) is substantially enhanced
compared to its zero-temperature value. As a consequence, a larger number $%
g_{\ast }$ than obtained in the SM calculation follows. To determine $%
g_{\ast }$ from the observed primordial abundances of light elements one
needs to perform the detailed simulation invoking the above dynamics. This
is beyond the scope of the present paper \bigskip \footnote{%
Ideal testing grounds for the postulate SU(2)$_{W}=$SU(2)$_{e}$ subject to
the nonperturbative approach of \cite{Hofmann2005} are the physics of the
solar core and the central plasma region of a state-of-the-art tokamak: The
constant flux of $10^{43}$ protons per year \cite{Manuel2004}, the solar
wind, badly violates electric charge conservation which, according to the
SM, should hold at the temperatures prevailing in the solar core. Moreover,
the onset of the Hagedorn transition from the confining to the preconfining
phase, which violates spatial homogeneity, possibly is detected by
micro-turbulences within the magnetically confined plasma of a tokamak at a
central temperature $\sim 40$\thinspace keV$\sim 1/10\,m_{e}$ \cite{ITER}.
For other indications that SU(2)$_{W}=$SU(2)$_{e}$ is in agreement with
experiment see \cite{Hofmann2005} and related references therein.}.

To summarize, as far as the first lepton family and its interactions is
concerned the electroweak sector of the SM is likely described by pure SU(2)$%
_{{\tiny \mbox{CMB}}}\times $SU(2)$_{e}$ dynamics where the gauge modes of
SU(2)$_{e}$ are very massive for $T<m_e$. In a similar way the doublet $(\nu
_{\mu },\mu )$ corresponds to the (stable) excitations of an SU(2)$_{\mu }$
pure gauge theory with $\Lambda _{\mu }\sim m_{\mu }\sim 200\,m_e$ (and also
to three very massive intermediary gauge modes $\Omega^{\pm},\Omega_0$ not
yet detected). The corresponding group structure is a direct product of
SU(2) Yang-Mills factors responsible for the existence of leptons and their
interactions: 
\begin{equation}  \label{prodga}
\mbox{SU(2)}_{{\tiny \mbox{CMB}}}\times \mbox{SU(2)}_{e}(=\mbox{SU(2)}_{W})
\times \mbox{SU(2)}_{\mu }\times \cdots
\end{equation}
with nontrivial mixing\footnote{%
As for the mixing a tacit assumption is that the above gauge symmetry is a
remnant of a breaking SU(N$\gg $1)$\rightarrow $SU(2)$_{{\tiny \mbox{CMB}}%
}\times $SU(2)$_{e}(=$SU(2)$_{W})\times $SU(2)$_{\mu }\times \cdots $ at
energies not too far below the Planck mass.}. In view of the above scenario
the electroweak sector of the SM emerges as a low-energy effective theory
being valid for momentum transfers ranging from zero up to values not much
larger than $m_{Z_0}\sim 91\,$GeV for an isolated vertex or for temperatures
not exceeding $m_e\sim 0.5\,$MeV.

\noindent\textit{(b)}$\gamma $\textit{-}$V^{\pm }$\textit{\ interaction.}

Let us now discuss why the $\gamma $ excitation of SU(2)$_{{\tiny \mbox{CMB}}%
}$ does practically not radiate off or create $V^{\pm }$ pairs, why $\gamma $
is is not created by the annihilation thereof and why there is practically
no scattering of $\gamma $ off of $V^{\pm }$.

In Sec.\,\ref{dp} we introduced the Higgs field $\varphi _{E},$ describing
the BPS saturated part of the ground-state dynamics in the deconfining
phase. In a physical (unitary-Coulomb) gauge, quantum fluctuations of
nontrivial and trivial topology are integrated out down to a resolution $%
|\varphi _{E}|$ in the effective theory. As a consequence, one has for the
off-shellness of residual quantum fluctuations 
\begin{equation}
|p^{2}-m^{2}|\leq |\varphi _{E}|^{2}  \label{offsh}
\end{equation}%
for the momentum transfer in a four vertex (deconfining phase)\footnote{%
We do not distinguish $s$, $t$, and $u$ channels here. See, however, \cite%
{Hofmann2006}.} 
\begin{equation}
|(p+q)^{2}|\leq |\varphi _{E}|^{2} \,.  \label{momtrans}
\end{equation}%
Notice that the constraints eqs.\,(\ref{offsh}) and (\ref{momtrans}) do not
apply close to the Hagedorn transition at $T_{c,M};$ in fact, in the
critical region preconfining phase $\leftrightarrow $ confining phase
thermal equilibrium breaks down: the 't Hooft loop undergoes rapid and local
phase changes violating spatial homogeneity \cite{Hofmann2005}. Thus a limit
on the maximal resolution, as it emerges in the thermalized situation ($T\ll
T_{c,M}$ and $T>T_{c,M}$), no longer exists close to the Hagedorn transition.

If it were not for the constraints eqs.\thinspace (\ref{offsh}) and (\ref%
{momtrans}) the effective theory would be strongly interacting, recall that $%
e\sim 8.89$ for $T\gg T_{c,E}$), and the postulate SU(2)$_{{\tiny \mbox{CMB}}}%
\overset{{\tiny \mbox{today}}}{=}$U(1)$_{Y}$ surely would not be viable. As $%
|\varphi _{E}|$ decays like $|\varphi _{E}|=\sqrt{\Lambda _{E}^{3}/2\pi T}$
the constraints eqs.\thinspace (\ref{offsh}) and (\ref{momtrans}) become
tighter and tighter with increasing temperature $T$. For example, the
modulus of the ratio of two-loop corrections to the one-loop result for the
thermodynamical pressure in the deconfining phase rapidly approaches $%
4\times 10^{-4}$ for $T\gg T_{c,E}$ and has a peak of $\sim 10^{-2}$ at $%
T\sim 3\,T_{c,E}$ \cite{SHG2006,HerbstHofmannRohrer2004}. Thus the tiny interactions
at high temperature can be absorbed into a tiny shift of the temperature in
a free-gas expression of the pressure for massless gauge modes, see the
Appendix.

There is, indeed, a regime $T\overset{\sim }{>}T_{{\tiny \mbox{CMB}}}$,
where a small fraction $\sim 10^{-3}$ of the $\gamma $ excitations is
converted into $V^{\pm }$ pairs and vice versa or where there is very mild
scattering of $\gamma $ off of $V^{\pm }$. While this is of (computable)
relevance for CMB physics at redshift $z\sim 2-10$ or so \cite%
{SHG2006,HerbstHofmannRohrer2004} there is no measurable effect in collider
experiments, atomic physics, and astrophysical systems \footnote{%
Radiowave propagation occurs on a thermalized CMB background with decoupled $%
V^{\pm }$ today. That is, the present Universe's thermalized ground state
does not allow for the creation of these particles as intermediary
fluctuations left alone their on-shell propagation. Exceptional
astrophysical systems could be the dilute, old, and cold clouds of atomic
hydrogen which are observed in between spiral arms of our galaxy, see \cite%
{SHG2006} and references therein.}.

\noindent \textit{(c) Charged currents.}

By virtue of the gauge structure proposed in eq.\,(\ref{prodga}) one would
expect the presence of additional charged-current interactions. Let us
discuss the prototype of such an interaction in the SM: the decay $\mu ^{\pm
}\rightarrow e^{\pm }+\bar{\nu}_{e}+\nu_{\mu}$. The immediate question is
why this decay, which is mediated by an intermediary $W^{\pm }$ in the SM,
is not enhanced by $V^{\pm }$ mediation through a nontrivial mixing of $%
W^{\pm }$ and $V^{\pm }$. On the scale of $m_{\mu }\sim 10^{12}\times \,T_{%
{\tiny \mbox{CMB}}}$ the $V^{\pm }$ modes are practically massless even if
we assume a decoupling mass $\sim 10^{5}\times \,T_{{\tiny \mbox{CMB}}}$ in
analogy to the experimentally accessible case of SU(2)$_{e}$. We have $%
|\varphi_{E}|\sim T_{{\tiny \mbox{CMB}}}$. Now the momentum transfer in the
decay is comparable to $m_{\mu }$ and thus the condition in eq.\thinspace (%
\ref{offsh}) would be badly violated for a $V^{\pm }$ intermediary
fluctuation. Thus mediation of the decay by $V^{\pm }$ is strictly
forbidden. Mediation by an $W^{\pm }$ intermediary fluctuation is, however,
allowed through mixing SU(2)$_{{\tiny \mbox{CMB}}}\leftrightarrow$SU(2)$_e$
although this particle is also far away from its mass shell: in contrast to $%
V^{\pm }$ $W^{\pm }$ is virtually excited \textsl{across} the Hagedorn phase
boundary of SU(2)$_{e}$ where the constraints in eqs.\thinspace (\ref{offsh}%
) and (\ref{momtrans}) do not apply. The exclusion of $V^{\pm }$ mediation
in $\mu ^{\pm }$ decay represents all other charged-current processes with a
momentum transfer exceeding $m_V^{\pm}\sim 10\,$eV. The discussion in \textsl{%
(b)} and \textsl{(c)} is important to not contradict the neutrino luminosity
measured in the SN 1987A cooling pulse\footnote{%
Spherical 2D models of neutrino-driven heating of the stellar plasma around
the nascent neutron star do not generate the observed supernova explosions 
\cite{Janka:2000ce,Burrows:2002bz}. Although explosions may arise from
hydrodynamical instabilities induced by medium anisotropies \cite%
{Scheck:2006rw} a possibility for a more efficient neutrino heating of the
stellar plasma may be an enhanced Fermi coupling $G_F$ as compared to the
zero-temperature and density value.}, see for example \cite{Choi1990}, and
to be below the imposed experimental cuts for missing momenta in collider
experiments.

Let us remark that on the level of eq.\,(\ref{prodga}), that is, resolving
the local SM vertex, the decaying soliton $\mu ^{\pm }$ first must
(nonlocally) couple to the soliton $\nu _{\mu }$ and to $\Omega ^{\pm }$
which subsequently rotates into $W^{\pm }$. The latter (nonlocally) couples
to the solitons $\nu _{e}$ and $e^{\pm }$. On the effective level of the SM
only a $W^{\pm }$ mediation appears with \textsl{local} coupling to $\mu
^{\pm }$, $\nu _{\mu }$, $\nu _{e}$, and $e^{\pm }$ which are all treated as
point particles. The SM vertex for charged currents follows from an
(effective) SU(2)$_{W}$ gauge principle. Obviously, its derivation in terms
of complex dynamics governed by eq.\,(\ref{prodga}) is an extremely
complicated task, see for example \cite{forkel2005}. It may or may not be
accomplished in the future. Seen in this light, the SM is an effective (and
ingenious) quantum field theoretic set up describing the \textsl{interactions%
} between \textsl{postulated} point particles of given (effective) gauge
charge. The gauge structure proposed in eq.\,(\ref{prodga}) facilitates a
deeper understanding of gauge-symmetry breaking, of the ground-state
structure of our Universe, of zero-temperature particle properties such as
the classical magnetic moment and the classical selfenergy (mass of the
electron, ...) and of the high-temperature behavior of particle physics.

\noindent\textit{(d) $\gamma$ interaction with charged leptons and neutral
currents.}

There is an important difference with the charged-current case. Namely, 
\textsl{electrically} charged (with respect to the defining fields of
SU(2)), far-off-shell $W^{\pm }$ bosons \textsl{induce} \textsl{magnetically}
charged monopoles\footnote{%
By `induce' we mean that the interaction between magnetic and electric
charges necessarily is highly nonlocal \cite{nonem}.}, represented by the
selfintersection region of a center-vortex loop, and vice versa while there
is a \textsl{coupling} of the dual gauge boson to the \textsl{magnetically}
charged monopoles. By virtue of eq.\thinspace (\ref{offsh}) the excitation $%
\gamma $ of SU(2)$_{{\tiny \mbox{CMB}}}$ is only allowed a maximal
off-shellness comparable to $T_{{\tiny \mbox{CMB}}}$. This means that by
itself it cannot mediate electromagnetic interactions with momentum transfer
on atomic physics scales or on even higher intrinsic scales. It could not do
so anyway in the absence of a mixing between the propagating excitations of
SU(2)$_{{\tiny \mbox{CMB}}}$ and SU(2)$_{e}$, SU(2)$_{\mu }$, $\cdots $
because the $\gamma $ excitation simply would not `see' the charged leptons.
Since such a (universal) mixing exists a sufficiently off-shell $\gamma $
mode never is emitted by a charged lepton but rather the associated dual
gauge mode $Z_{0}$, $\Omega_0$, $\cdots $ of SU(2)$_{e}$, SU(2)$_{\mu }$, $%
\cdots $. In contrast to $\gamma $ the latter are allowed to be off-shell
across their respective Hagedorn boundaries. The $Z_{0}$, $\Omega_0$, $%
\cdots $, in turn, couple to their charged leptons with a large magnetic
coupling $g_{\mbox\tiny{dec}}\sim 10^{5}$. In contrast to the
charged-current process associated with a parametric suppression $\frac{p^{2}%
}{m_{W}^{2}}$ ($p$ being the transferred momentum) the coupling of $Z_{0}$, $%
\Omega_0$, $\cdots $ to the associated charged leptons leads to an
enhancement $\frac{(g_{\mbox\tiny{dec}}p)^{2}}{m_{Z_{0}}^{2}}$ explaining
why electromagnetic interactions are so much stronger than weak
interactions. Again, the local SM vertex between a charged lepton and a
massless photon, determined by a universal (effective) U(1) gauge symmetry,
is an extremely efficient and successful description and very hard to be
derived from the underlying gauge dynamics with nonlocal interactions
subject to SU(2)$_{{\tiny \mbox{CMB}}}$, SU(2)$_{e}$, SU(2)$_{\mu }$, $%
\cdots $. One of the advantages of the latter description is, however, a
deeper grasp of the particle-wave duality of the photon: Let $\gamma $ be on
shell, thus propagating as a wave over large distances. Whenever $\gamma $
approaches a charged lepton it rotates into the associated massive $Z_{0}$, $%
\Omega_0$, $\cdots $ excitation to interact with the charge. This process
changes the wave into a massive particle transferring its momentum to the
lepton in the subsequent collision.

\section{Cosmological evolution from $z_{dec}=1089$ to $z=0$}

We consider a spatially flat Universe whose expansion is sourced by baryonic
and dark, pressureless matter ($M$), a homogeneous axion field $\phi $ and
SU(2)$_{{\tiny \mbox{CMB}}}$ Yang-Mills thermodynamics. The evolution of the
scale parameter $a=a(t)$ is determined by the Friedman equation%
\begin{equation}
H(t)^{2}=\left( \overset{\cdot }{a}/a\right) ^{2}=\frac{8\pi }{3}G\left(
\rho _{M}+\rho _{\phi }+\rho _{{\tiny \mbox{CMB}}}\right)  \label{h}
\end{equation}%
where $G\equiv \frac{1}{M_{P}^{2}}$ and $M_{P}\equiv 1.2209\times 10^{19}\,$%
GeV. We are only interested in the evolution after CMB decoupling, i.e. for $%
z\leq z_{dec}=1089$. Within this range the contribution of neutrinos can be
neglected. Each of the contributions to the right-hand side of eq.\thinspace
(\ref{h}) are associated with separately conserved cosmological fluids as
long as $z\geq 0$: 
\begin{equation}
d\left( \rho _{i}a^{3}\right) =-p_{i}d(a^{3}),\text{ }\ \ \ (i=M,\text{ }%
\phi ,\text{ }\mbox{CMB})\,.  \label{rho}
\end{equation}%
Since $p_{M}=0$ we have $\rho _{M}(a)=\rho _{M}(a_{0})\cdot (a_{0}/a)^{3}$
where $t_{0}$ is the present age of the universe (to be calculated) and $%
a_{0}\equiv a(t_{0})$. In terms of the critical density $\rho
_{c}=3\,H(t_{0})^{2}/8\pi G=4.08\times 10^{11}$ eV$^{4}$ the measured matter
contribution reads \cite{pdgealtri}: 
\begin{equation}
\Omega _{M}=\frac{\rho _{M}}{\rho _{c}}(a_{0})=\Omega _{{\tiny %
\mbox{Dark-Matter}}}+\Omega _{{\tiny \mbox{Baryon}}}=0.27\pm 0.04\,.
\label{omegamatter}
\end{equation}%
By virtue of eq.\thinspace (\ref{rho}) (for the equation of state $p_{{\tiny %
\mbox{CMB}}}=p_{{\tiny \mbox{CMB}}}(\rho _{{\tiny \mbox{CMB}}})$ see
Appendix) the dependence $\rho _{{\tiny \mbox{CMB}}}=\rho _{{\tiny \mbox{CMB}%
}}(a)$ is calculated numerically. Notice, however, that at $t=t_{dec},$ the
contribution of SU(2)$_{{\tiny \mbox{CMB}}}$ to the critical energy density
is about 10\% and decreases very rapidly for $t>t_{dec}$: Although not
directly affecting the evolution of the Universe, the presence of SU(2)$_{%
{\tiny \mbox{CMB}}}$ is imprinted in the potential for a Planck-scale axion
eq.\thinspace (\ref{pot}). We rewrite this potential as follows:%
\begin{equation}
V(\phi )=(\lambda \cdot \Lambda _{{\tiny \mbox{CMB}}})^{4}\left[ 1-\cos
\left( \frac{\phi }{F}\right) \right] \,.  \label{pot2}
\end{equation}%
The dimensionless quantity $\lambda $ parameterizes the uncertainty in the
coupling of the topological defects of SU(2)$_{{\tiny \mbox{CMB}}}$ to the
axion. The value of $\lambda $ is expected to lie within $O(10^{-1})$ to $%
O(10^{1})$ \cite{PecceiQuinn1977}. In our calculation we adjust $\lambda $
such that the measured value of dark energy density is reproduced today \cite%
{pdgealtri}: 
\begin{equation}
\Omega _{\phi }=\frac{\rho _{\phi }}{\rho _{c}}(a_{0})=1-\Omega _{M}=0.73\pm
0.04\,.  \label{omegaaction}
\end{equation}%
The axion energy density $\rho _{\phi }$ and the pressure $p_{\phi }$ are
given as%
\begin{equation}
\rho _{\phi }=\frac{1}{2}\overset{\cdot }{\phi }^{2}+V(\phi )\,,\ \ \ \ \ \
\ \ p_{\phi }=\frac{1}{2}\overset{\cdot }{\phi }^{2}-V(\phi )\,.
\label{rhopia}
\end{equation}%
>From (\ref{rho}) and (\ref{rhopia}) the equation of motion for $\phi $
follows:%
\begin{equation}
\overset{\cdot \cdot }{\phi }+3H\overset{\cdot }{\phi }+V^{\prime }(\phi
)=0\,  \label{phi}
\end{equation}%
where $V^{\prime }\equiv dV/d\phi $. The term $3H\overset{\cdot }{\phi }$
represents the cosmological \textquotedblleft friction\textquotedblright .

The origin of the field $\phi $ is due to the axial anomaly starting to be
operative before inflation. In \cite{frieman1995} it was concluded that the
CMB-constraints on $\phi$-induced adiabatic density perturbations be such
that the inflationary Hubble parameter is smaller than $10^{13}$\,GeV. This
entails that the scale $F$ be larger than $10^{18}$\,GeV$\simeq 0.1\,M_{P}$.
Moreover, a quantum field theoretic description in (3+1) dimensions, which
underlies the axial anomaly, likely is meaningful only below the Planck
mass. Thus it is natural to suppose that $F\sim M_{P}.$

The classical field $\phi$, representing a condensate of axion particles
being generated at $T\sim M_P$, is surely fixed to its starting value $%
\phi_{in}\sim F$ all the way down to CMB decoupling because of the large
cosmological ``friction''. This implies the following initial conditions at
decoupling: 
\begin{equation}  \label{indec}
\phi_{in}=\phi(t_{dec})\sim F\,,\ \ \ \ \ \dot\phi(t_{dec})=0\,.
\end{equation}
We first consider $0\le \phi _{in}\le \pi \frac{F}{2}$, i.e. a range for
which the curvature of the potential is positive. Let us now discuss the
conditions under which the axion field behaves like a cosmological constant
at present, that is, $\phi$ did not roll down its potential until now. This
happens if \footnote{$H(t)$ is a monotonically decreasing function, that is,
if the condition $3H(t_{0})\gg 2m_{\phi }$ is satisfied at $t_{0},$ it also
holds for $t_{dec}\le t\le t_0$.} 
\begin{equation}  \label{cond}
3H(t_0)\gg 2m_{\phi}\,
\end{equation}
where $m_{\phi }\equiv(\lambda \cdot \Lambda _{{\tiny \mbox{CMB}}})^{2}/F.$

By using eq.\thinspace (\ref{h}) and $V(\phi )\simeq \frac{1}{2}\,m_{\phi
}^{2}\phi ^{2}$ and neglecting the small direct contribution of SU(2)$_{%
{\tiny \mbox{CMB}}}$, we have 
\begin{equation}
H(t_{0})^{2}=\frac{4}{3}\,\pi \,G\,m_{\phi }^{2}\phi _{in}^{2}\left( 1+\frac{%
\Omega _{M}}{\Omega _{\phi }}\right) \,.  \label{hes}
\end{equation}%
Rewriting the condition (\ref{cond}) by using eq.\thinspace (\ref{hes}), we
derive: 
\begin{equation}
\frac{\phi _{in}}{M_{P}}\gg \frac{1}{\sqrt{3\pi (1+\Omega _{M}/\Omega _{\phi
})}}\simeq 0.278.  \label{ll}
\end{equation}%
Even for $\phi _{in}/M_{P}\gtrsim 0.278$ slowly rolling solutions compatible
with today's dark energy are numerically found, see discussion below. For $%
\phi _{in}/M_{P}\lesssim 0.278$ the parameter $\lambda $ needs to assume
unnaturally large values for the axion to generate today's value of dark
energy density. Moreover, the axion would undergo many oscillations until
today and thus would behave more like pressureless matter than dark energy.

For $0\leq \phi _{in}\leq \pi \frac{F}{2}$ to be meaningful when compared to
the constrain of eq. (\ref{ll}) one needs $F/M_{P}>0.177$. This is close to
the lower bound $F/M_{P}>0.1M_{P}$ arising from the consideration on
adiabatic density perturbations in \cite{frieman1995}. 
\begin{figure}[tbp]
\begin{center}
\leavevmode
\leavevmode
\vspace{4.3cm} \includegraphics{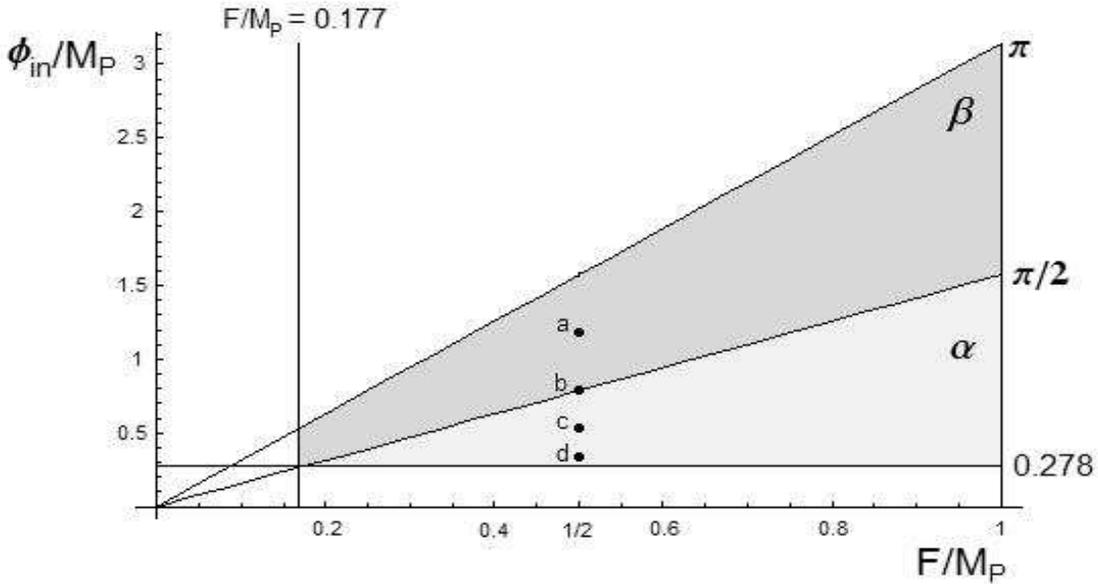}
\end{center}
\caption{{Admissible range of the quantities $\protect\phi _{in}/M_{P}$ and $%
F/M_{P}$ for dark-energy like axion-field solutions today. The triangular
area $\protect\alpha $ ($\protect\beta $) corresponds to $\protect\phi %
_{in}/M_{P}$ being below (above) the inflexion point of $V(\protect\phi )$.
The horizontal line $\protect\phi _{in}/M_{P}=0.278$ indicates a rapid
cross-over from slowly rolling to oscillating solutions.}}
\label{Fig-2}
\end{figure}
In Fig.\thinspace \ref{Fig-2} admissible ranges for the initial conditions
at $t=t_{dec}$ are shown. The triangular area $\alpha $ represents the
allowed parameter range for a slowly rolling field at present. The
horizontal line $\phi _{in}/M_{P}=0.278$ indicates a rapid crossover from
dark-energy-like (above) to oscillating (below) solutions. The allowed range
is enlarged by including the trapezoidal area $\beta $ corresponding to a
negative curvature of the potential\footnote{%
If the field does not roll at $\phi _{in}=\pi F/2$ (inflexion point) then it
also does not roll for $\pi F/2\leq \phi _{in}\leq \pi F$.}. Notice that for 
$\pi F/2\leq \phi _{in}\leq \pi F$ there are slowly rolling solutions with
the needed amount of present dark energy also for $F/M_{P}<0.177$. However,
for a decreasing value of $F/M_{P}$ we observe that $\phi _{in}$ needs to be
closer to the maximum $\pi F$ which is somewhat of a fine-tuned situation 
\cite{kaloper2005}. We thus pick representative initial conditions as
depicted in Fig\thinspace \ref{Fig-2}.

In Table 1 we present our numerical results obtained for initial values
corresponding to the points (a),(b),(c) and (d) in Fig.\,\ref{Fig-2}. The
values of the following quantities are determined: $\lambda$ such that $%
\Omega _{\phi }=0.73$ at present, the present age $t_{0}$ of the universe,
the present Zeldovich parameter for the axion fluid alone, $w_{\phi
}(t_{0})=(p_{\phi }/\rho _{\phi })_{t=t_{0}}$ (see eq.\,(\ref{rhopia})), and
for the entire Universe, $w_{tot}(t_{0})=(p_{tot}/\rho _{tot})_{t=t_{0}}$
and the value of redshift $z_{acc}$ corresponding to the transition between
decelerated and accelerated expansion. 

\begin{center}
\textbf{Table 1.} The values of selected cosmological parameters obtained
for variable \phantom{1} initial values at CMB decoupling keeping $%
F/M_{P}=0.5$ fixed.\vspace{0.2cm}\\[0pt]
\begin{tabular}{|l|l|l|l|l|l|l|}
\hline
points in Fig.\thinspace\ref{Fig-2} & $\phi _{in}/M_{P}$ & $\lambda $ & $%
t_{0}$\thinspace (Gy) & $w_{\phi }(t_{0})$ & $w_{tot}(t_{0})$ & $z_{acc}$ \\ 
\hline
(a) & $3\pi /4$ & $22.15$ & $13.65$ & $-0.97$ & $-0.71$ & $0.76$ \\ \hline
(b) & $\pi /4$ & $22.15$ & $13.65$ & $-0.97$ & $-0.71$ & $0.76$ \\ \hline
(c) & $\pi /6$ & $26.96$ & $13.56$ & $-0.91$ & $-0.66$ & $0.79$ \\ \hline
(d) & $0.328$ & $37.27$ & $13.08$ & $-0.61$ & $-0.44$ & $0.92$ \\ \hline
\end{tabular}
\end{center}

For the set of initial values (a),(b), and (c) the axion field does not roll
until $t_{0},$ as indicated by the quantity $w_{\phi }(t_{0})\simeq -1$. For
point (d) $\phi _{in}/M_{P}$ is just above the threshold in (\ref{ll})
causing the field $\phi$ to roll at present: $w_{\phi }(t_{0})=-0.61$.
According to ref.\,\cite{pdgealtri} $w_{\phi }(t_{0})=-0.61$ is already
inconsistent with observation ($w_{\phi }(t_{0})<-0.78$ at 95\% C.L.).
Decreasing $\phi_{in}/M_{P}$ further one rapidly runs into the regime $%
w_{tot}(t_{0})>-1/3$ where the present Universe does not accelerate. The
values of $z_{acc}$ obtained for (a), (b), and (c) are in approximate
agreement with $z_{acc}=0.75$ obtained for a standard $\Lambda$CDM model.
Moving $\frac{F}{M_P}$ within the allowed range $\alpha\cup\beta$ at fixed
values of $\phi_{in}/M_{P}$, see Fig.\,\ref{Fig-2}, the values of the
cosmological parameters in Tab.\,1 are almost unaffected.

Due to the dynamical nature of dark energy in our model the Universe will
not run into pure de Sitter expansion in the future as it does for the $%
\Lambda$CDM model but rather epochs of accelerated and decelerated
expansions will alternate: $z_{acc}$ corresponds to the first of many more
future turning points ($\overset{\cdot \cdot }{a}=0$). This, however,
presumes that the axion-SU(2)$_{{\tiny \mbox{CMB}}}$ coupling will remain
unaffected by the future evolution.

\section{A massive photon in the future\label{masspho}}

Here we consider the future evolution up to the point where SU(2)$_{{\tiny %
\mbox{CMB}}}$ undergoes the transition to its preconfining phase. For
simplicity we assume that the present age of the Universe $t_{0}$ is given
by the time when $T_{0}=T_{{\tiny \mbox{CMB}}}$ is first reached. Because of
the discontinuity of the energy density (Fig.\,\ref{Fig1})%
\begin{equation}
\frac{\rho _{c,M}}{T_{{\tiny \mbox{CMB}}}^{4}}-\frac{\rho _{c,E}}{T_{{\tiny %
\mbox{CMB}}}^{4}}=\frac{4}{3}\frac{\pi ^{2}}{30}
\end{equation}%
the system cannot jump into the preconfining (magnetic) phase where the
photon possesses a mass and thus an additional polarization because of
condensed monopoles in the ground state. The energy gap $\frac{4}{3}\frac{%
\pi ^{2}}{30}$ is the sum of the energy gap of the photon gas ($=\frac{\pi
^{2}}{30}$) and of the ground state ($=\frac{1}{3}\frac{\pi ^{2}}{30}$); we
refer to the Appendix for the details. Therefore, the system remains in the
deconfining (electric) phase in a supercooled state (with its ground state
still being an ensemble of interacting calorons instead of monopoles) so
long as the energy density of the electric phase $\rho _{E}$ is smaller than
the energy density of the magnetic phase $\rho _{M}$ ($\rho _{E}<\rho _{M}$%
). At a certain value of temperature, $\lambda _{\ast ,E}=2\pi T_{\ast
}/\Lambda _{E}<\lambda _{c,E}=2\pi T_{c,E}/\Lambda _{E}$, equality $\rho
_{E}=\rho _{M}$ takes place. At this point the condensation of monopoles
occurs and the photon becomes massive (for $\lambda _{\ast ,E}\leq \lambda
_{E}\leq \lambda _{c,E}$ monopoles are not sufficiently liberated by the
associated large-holonomy calorons to facilitate unlimited mobility). 
\begin{figure}[tbp]
\begin{center}
\leavevmode
\leavevmode
\vspace{5.3cm} \includegraphics{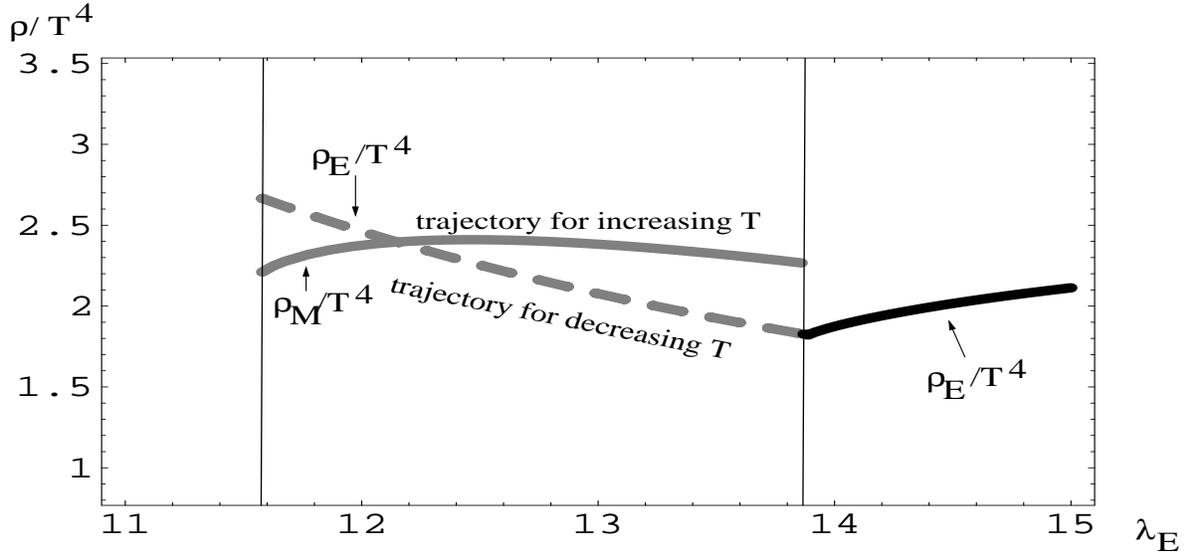}
\end{center}
\caption{{The situation for the (dimensionless) energy density in the
critical region: the dashed line represents the continuation of the energy
density of the deconfining phase (solid grey line) for $T<T_{c,E}$
(supercooled state, realized for decreasing temperature, 
$m_\gamma =0$). The solid grey line depicts the energy density in the
preconfining phase (realized for increasing temperature, $m_\gamma>0$). At the intersection point $\protect\lambda _{E}=\protect%
\lambda _{\ast ,E}$ a phase transition from the supercooled deconfining to
the preconfining dynamics occurs.}}
\label{Fig-3}
\end{figure}
The situation is depicted in Fig.\,\ref{Fig-3}, where the dashed line represents the
continuation of the (dimensionless) energy density $\rho _{E}/\Lambda
_{E}^{4}$ for $\lambda _{E}<\lambda _{c,E}$. The corresponding analytical
expressions are given in the Appendix \cite{Hofmann2005}. The intersection $%
\rho _{E}/T^{4}=\rho _{M}/T^{4}$ occurs at $\lambda _{\ast ,E}=12.15<\lambda
_{c,E}=13.87$ (i.e. $T_{\ast }/T_{c,E}=0.88$) (see Fig.\thinspace\ref{Fig-3}%
, for technical details see the Appendix). Driven by cosmological expansion,
which essentially is sourced by dark matter and the axion field, the SU(2)$_{%
{\tiny \mbox{CMB}}}$ thermodynamics evolves into a supercooled state
(deconfining phase) according to eq. (\ref{rho}) ($i=$ CMB) down to the
point where the density $\rho _{\ast }$ (i.e. the transition temperature $%
T_{\ast }$) is reached. The numerical result for the scale factor at $%
T=T_{\ast }$ is $a(t_{\ast })/a_{0}=1.17.$ At this point the photon acquires
a Meissner mass. Notice that according to eq.\thinspace (\ref{lag}) the
anomaly-mediated decay width $\Gamma _{\phi \rightarrow 2\gamma }$ of the
axion into two photons is much smaller than the present Hubble parameter $%
H_{0}$:%
\begin{equation}
\Gamma _{\phi \rightarrow 2\gamma }<\left( \frac{m_{\phi }}{F}\right)
^{2}\,m_{\phi }\,\,\sim 10^{-155}\text{ eV}\lll H_{0}\sim 10^{-33}\text{ eV.}
\end{equation}%
Thus it is justified to treat the axion as a coherent field for any
practical purpose and to consider the axion and the SU(2)$_{{\tiny \mbox{CMB}%
}}$ fluids to be separately conserved as in eq.\thinspace (\ref{rho}).

The numerical value of the time interval $\Delta t_{m_{\gamma
}=0}=t_{*,E}-t_{0}$ follows from future cosmology according to eq. (\ref{h}%
). For the sets of initial values (a)--(d) in Tab.\thinspace 1 we obtain the
following numbers: 
\begin{eqnarray}
(a) &&\ \ \Delta t_{m_{\gamma }=0}=2.20\,\mbox{Gy}\,,\ \ \ \ \ (b)\ \ \Delta
t_{m_{\gamma }=0}=2.20\,\mbox{Gy}\,,  \notag  \label{deltat} \\
(c) &&\ \ \Delta t_{m_{\gamma }=0}=2.22\,\mbox{Gy}\,,\ \ \ \ \ (d)\ \ \Delta
t_{m_{\gamma }=0}=2.29\,\mbox{Gy}\,.
\end{eqnarray}%
The value $\Delta t_{m_{\gamma }=0}\simeq $ 2.2\,Gy depends only weakly on the
chosen parameter set. The error in determining the quantity $\Delta
t_{m_{\gamma }=0}$ is dominated by the observational uncertainty for the
present Hubble parameter $H_0$. According to \cite{pdgealtri} we have $\frac{%
\delta H(t_{0})}{H(t_{0})}\sim \pm 0.056$. We have run our simulations with
the upper (lower) limit for the error range. This generates a decrease
(increase) for $\Delta t_{m_{\gamma }=0}$ of about 0.15\,Gy.

Throughout the work we assumed that the temperature $T_{c,E}=T_{{\tiny %
\mbox{CMB}}}$ is reached today. However, the photon is massless and
unscreened in the entire range $T_{\ast }$ $\leq T\leq T_{c,E}$. Therefore
the present CMB-temperature could also be below $T_{c,E}$. As a consequence,
the quantity $\Delta t_{m_{\gamma }=0}$ represents an upper bound for the
time interval between the present and the occurrence of the phase transition.

The existence of extra-galactic magnetic fields \cite{Dai2002,Bagchi2002}
could indeed signal the onset of a superconducting vacuum. Possibly, a
quantitative analysis would rely on tunneling effects connecting the two
trajectories in Fig.\,\ref{Fig-3}. Such an unconventional interpretation of
extra-galactic magnetic fields needs future investigation.

\section{Summary and Conclusions}

We have elaborated on the idea that the present density of dark energy
arises from an ultra-light axion field with Peccei-Quinn scale comparable to 
$M_{P}$ \cite{frieman1995}. More precisely, we have linked the normalization 
$\mu ^{4}$ of the axion potential in eq.\thinspace (\ref{pot}) to the
existence of an SU(2) Yang-Mills theory of scale $\Lambda _{{\tiny \mbox{CMB}%
}}\sim \mu $ comparable to the temperature $T_{{\tiny \mbox{CMB}}}$ of the
present cosmic microwave background. Such an assertion has its justification
in nonperturbative results obtained recently for SU(2) Yang-Mills
thermodynamics \cite{Hofmann2005}. As a result, we have obtained an upper
bound $\Delta t_{m_{\gamma }=0}\simeq 2$\,Gy for the length of the time
interval from the present to the phase transition where the photon acquires
a Meissner mass.

Throughout our work we have assumed a cold dark matter component $\rho
_{M}=0.23\,\rho _{c}$ of unknown origin at present. (A possibility would be
that $\rho _{M}$ arises due to the decay of one or more oscillating,
coherent axion fields $\phi_i$ with $F_i\ll M_P$ into their particles at
earlier epochs.) A more unified but also more speculative picture would
arise if today's rolling axion field would describe both dark matter and
dark energy, see \cite{Pad2002} and refs. therein. On the one hand,
according to our simulations (performed with a canonical kinetic term) such
a scenario would imply an age of the Universe of about 20\thinspace Gy as
opposed to 13.7\thinspace Gy with conventional cold dark matter. Also one
would obtain $z_{acc}\sim 3$ as opposed to $z_{acc}\sim 0.75$, possibly
endangering structure formation. On the other hand, structure formation and
the flattening of the rotation curves of galaxies would need an explanation
in terms of ripples and lumps of a coherent axion field \cite{Wetterich2001}%
. Moreover, the relation between luminosity distance and redshift as
observed from SNe Ia standard candles would have to be postdicted with a
pressureless contribution to the Hubble parameter that acquired nominal
strength only very recently. The future will tell (gravitational lensing
signatures for galaxies, theoretical results on the stability of the system
axion-lump plus baryonic matter plus gravity) whether such a possibility is
viable.

For completeness we have investigated how the latter scenario affects our
estimate $\Delta t_{m_{\gamma }=0}$. By defining the quantity $\eta $
through $\eta =-p_{\phi }$ and $\rho _{\phi }=\dot{\phi}^{2}+\eta $ the
axion fluid can be split into a component with $\rho _{\Lambda }=\eta $
(with $w_{\Lambda }=-1$) and a component $\rho _{DM}=\dot{\phi}^{2}$ (with $%
w_{DM}=0$). Notice that the so defined components are not separately
conserved. The task is to uniquely fix $\phi _{in}$ and $\lambda $ in
eq.\thinspace (\ref{pot2}) such that $\Omega _{\phi }=0.96$ today (with $%
\Omega _{{\tiny \mbox{Baryon}}}=0.04)$) and such that $\Omega _{\Lambda
}=0.73$ and $\Omega _{DM}=0.23$. Using $\frac{F}{M_{P}}=0.5$ we obtain $%
\frac{\phi _{in}}{M_{P}}=0.53$ and $\lambda =31.9$. This yields $\Delta
t_{m_{\gamma }=0}=2.21\,$Gy. Thus our estimate $\Delta t_{m_{\gamma }=0}$ is
rather model independent.

Finally, let us make a few comments concerning future activity. The
postulate SU(2)$_{{\tiny \mbox{CMB}}}\overset{{\tiny \mbox{today}}}{=}$U(1)$%
_{Y}$ entails consequences for the CMB map of fluctuations in temperature
and in electric/magnetic field polarization at large angles \cite%
{SHG2006,HerbstHofmannRohrer2004}. To analyze these effects more quantitatively
needs precise information on the underlying cosmology; this basic step was
addressed in the present work.

For the viability of the postulate SU(2)$_{{\tiny \mbox{CMB}}}\overset{%
{\tiny \mbox{today}}}{=}$U(1)$_{Y}$ an alternative interpretation of
electroweak SM physics in terms of underlying, nonperturbative and pure
Yang-Mills dynamics is necessary. Although we have checked a few
experimental benchmarks on this scenario further theoretical work surely is
needed.

\section*{Acknowledgments}

The authors would like to thank Dirk Rischke for stimulating conversations.
F. G. acknowledges financial support by the Virtual Institute VH-VI-041
"Dense Hadronic Matter \& QCD Phase Transitions" of the Helmholtz
Association.

\appendix

\section{Expressions for the energy density and the pressure}

We start from the effective Lagrangians in the deconfining and preconfining
phases as described in \cite{Hofmann2005} and we briefly derive in a
self-contained way the corresponding 1-loop thermodynamical quantities.
Higher order loop corrections turn out to be of the order 0.1\% \cite%
{HerbstHofmannRohrer2004}, thus are irrelevant for our cosmological application.\vspace{0.1cm}\\  
\noindent{\bf Deconfining (electric) phase:}\\ 
(Occurs for $T>T_{c,E}=\Lambda _{E}\lambda _{c,E}/2\pi$, $\lambda
_{c,E}=13.87$, $\Lambda _{E}$ the Yang-Mills scale in electric phase).\newline
The effective Lagrangian for the description of SU(2)-Yang-Mills
thermodynamics in the deconfining phase and in the unitary gauge reads \cite%
{Hofmann2005}:
\begin{equation}
\mathcal{L}_{{\tiny \mbox{dec-eff}}}^{u.g.}=\frac{1}{4}\left( G_{E}^{a,\mu
\nu }[a_{\mu }]\right) ^{2}+2e(T)^{2}\left\vert \varphi _{E}\right\vert
^{2}\left( \left( a_{\mu }^{(1)}\right) ^{2}+\left( a_{\mu }^{(2)}\right)
^{2}\right) +\frac{2\Lambda _{E}^{6}}{\left\vert \varphi _{E}\right\vert ^{2}%
}\,,  \label{lageffdec}
\end{equation}%
where $G_{E,\mu \nu }^{a}=\partial _{\mu }(a_{\nu }^{a})-\partial _{\nu
}(a_{\mu }^{a})+e\varepsilon ^{abc}a_{\mu }^{b}a_{\nu }^{c}$ is the SU(2)
stress-energy tensor for the topologically trivial fluctuations $a_{\mu }^{a}$
(with effective coupling $e=e(T)$) and the adjoint scalar background-field $%
\varphi _{E}$ embodies the spatial coarse graining of caloron and
anticaloron field configurations (see \cite{Hofmann2005} for a microscopic
derivation, see \cite{garfield} for a macroscopic one). The quantum
fluctuations $a_{\mu }^{(1,2)},$ in our work identified by $V^{\pm },$ are
massive, while the gauge mode $a_{\mu }^{(3)},$ here the photon, stays
massless (spontaneous symmetry breaking $SU(2)\rightarrow U(1)$). The mass
of $V^{\pm }$ reads explicitly (see (\ref{lageffdec})): 
\begin{equation}
m=m_{V^{+}}=m_{V^{-}}=2e(T)\left\vert \varphi _{E}\right\vert ;\text{ }%
\left\vert \varphi _{E}\right\vert =\sqrt{\frac{\Lambda _{E}^{3}}{2\pi T}}\,.
\end{equation}%
At this stage the effective running coupling $e=e(T)$ is not yet known. Its
behavior is determined by imposing the thermodynamical self-consistency, see
below.

The energy density and the pressure are the sum of three terms, 
\begin{equation}
\rho _{E}=\rho _{E,\gamma }+\rho _{E,V^{\pm }}+\rho _{E,gs},\text{ }%
p_{E}=p_{E,\gamma }+p_{E,V^{\pm }}+p_{E,gs}\,,
\end{equation}%
corresponding to the contributions of the massless gauge mode $\gamma $, the
two massive gauge modes $V^{\pm }$ and the ground state, respectively. The 
1-loop expressions are easily obtained from the effective Lagrangian (\ref%
{lageffdec}):%
\begin{equation}
\rho _{E,\gamma }=2\frac{\pi ^{2}}{30}\,T^{4},\text{ }\rho _{E,V^{\pm
}}=6\int_{0}^{\infty }\frac{k^{2}dk}{2\pi ^{2}}\frac{\sqrt{m^{2}+k^{2}}}{%
\exp (\frac{\sqrt{m^{2}+k^{2}}}{T})-1},\text{ }\rho _{E,gs}=4\pi \Lambda
_{E}^{3}T\,
\end{equation}%
\begin{equation}
p_{E,\gamma }=2\frac{\pi ^{2}}{90}T^{4},\text{ }p_{E,V^{\pm
}}=-6\,T\int_{0}^{\infty }\frac{k^{2}dk}{2\pi ^{2}}\ln \left( 1-e^{-\frac{%
\sqrt{m^{2}+k^{2}}}{T}}\right) ,\text{ }p_{E,gs}=-\rho _{E,gs}\,.
\end{equation}

We first rewrite the system in terms of dimensionless quantities:%
\begin{equation}
\overline{\rho }_{E}=\frac{\rho _{E}}{T^{4}},\text{ }\overline{p}_{E}=\frac{%
p_{E}}{T^{4}},\text{ }\lambda =\lambda _{E}=\frac{2\pi T}{\Lambda _{E}},%
\text{ }a(\lambda )=\frac{m(T)}{T}=2\frac{e(T)}{T}\left\vert \varphi
_{E}\right\vert
\end{equation}%
where the function $a=a(\lambda )$ has been introduced for later use.

The dimensionless density and pressure, expressed as functions of the
dimensionless temperature $\lambda $, read:%
\begin{equation}
\overline{\rho }_{E,\gamma }=2\frac{\pi ^{2}}{30},\text{ }\overline{\rho }%
_{E,V^{\pm }}=\frac{3}{\pi ^{2}}\int_{0}^{\infty }dx\frac{x^{2}\sqrt{%
x^{2}+a^{2}}}{e^{\sqrt{x^{2}+a^{2}}}-1},\text{ }\overline{\rho }_{E,gs}=%
\frac{2(2\pi )^{4}}{\lambda ^{3}}\,.  \label{rhoad}
\end{equation}%
\begin{equation}
\overline{p}_{E,\gamma }=2\frac{\pi ^{2}}{90},\text{ }\overline{p}_{E,V^{\pm
}}=-\frac{3}{\pi ^{2}}\int_{0}^{\infty }x^{2}dx\ln \left( 1-e^{-\sqrt{%
x^{2}+a^{2}}}\right) ,\text{ }\overline{p}_{E,gs}=-\text{ }\overline{\rho }%
_{E,gs}\,.  \label{pad}
\end{equation}

We impose the validity of the thermodynamical Legendre transformation 
\begin{equation}
\rho =T\frac{dP}{dT}-P\iff \overline{\rho }=\lambda \frac{d\overline{p}}{%
d\lambda }+3\overline{p}  \label{tsc}
\end{equation}%
(expressed with both dimensional and dimensionless functions). By
substituting the expressions (\ref{rhoad}), (\ref{pad}) into (\ref{tsc}) we
determine the following differential equation for $a=a(\lambda )$:%
\begin{eqnarray}
1 &=&-\frac{6\lambda ^{3}}{(2\pi )^{6}}\left( \lambda \frac{da}{d\lambda }%
+a\right) aD(a),\text{ } \\
D(a) &=&\int_{0}^{\infty }dx\frac{x^{2}}{\sqrt{x^{2}+a^{2}}}\frac{1}{e^{%
\sqrt{x^{2}+a^{2}}}-1},\text{ }a(\lambda _{in})=0.
\end{eqnarray}%
For sufficiently large initial value $\lambda _{in}$ the solution for $%
a(\lambda )$ is independent on $\lambda _{in}$: a low-temperature attractor
with a logarithmic pole at $\lambda _{c,E}=$ $13.87$ is seen (infinite mass
for $V^{\pm }$ leading to their thermodynamical decoupling). The effective coupling is
given by $e=e(\lambda )=a(\lambda )\lambda ^{3/2}/4\pi $ and shows a
plateaux at $e\sim 8.89$, see Fig.\,\ref{Fig1B}.

Once the function $a=a(\lambda )$ is determined, density and pressure in the
deconfining phase are numerically obtained and are plotted in Fig.\,\ref{Fig1}.\vspace{0.1cm}\\ 
\noindent{\bf Preconfining (magnetic) phase:}\\ 
(Occurs for $11.57\,\Lambda _{E}/2\pi =T_{c,M}\leq T\leq
T_{c,E}=\Lambda _{E}\lambda _{c,E}/2\pi$).

The effective Lagrangian for the description of SU(2)-Yang-Mills
thermodynamics in the preconfining phase and in the unitary gauge reads \cite{Hofmann2005}:%
\begin{equation}
\mathcal{L}_{{\tiny \mbox{prec-eff}}}^{u.g.}=\frac{1}{4}(F_{E}^{\mu \nu
})^{2}+\frac{1}{2}g^{2}\left\vert \varphi _{M}\right\vert ^{2}(a_{\mu
}^{(3)})^{2}+\frac{2\Lambda _{M}^{6}}{\left\vert \varphi _{M}\right\vert ^{2}%
},\text{ }  \label{lageffmag}
\end{equation}%
where $F_{E}^{\mu \nu }=\partial _{\mu }a_{\nu }^{(3)}-\partial _{\nu
}a_{\mu }^{(3)}$ is the (dual) abelian stress-energy tensor, the complex scalar
field $\varphi _{M}$ describes the condensate of magnetic monopoles, and $\Lambda_M$ 
the Yang-Mills scale in the magnetic phase. The
remaining $U(1)$ symmetry of the deconfining phase is broken. The photon
acquires a temperature dependent mass:%
\begin{equation}
m_{\gamma }=g\left\vert \varphi _{M}\right\vert =b T\,,\text{ }\left\vert
\varphi _{M}\right\vert =\sqrt{\frac{\Lambda _{M}^{3}}{2\pi T}}
\end{equation}%
where the function $b\equiv b(\lambda )$ is introduced for later use.
The effective coupling $g\equiv g(\lambda)$ is not yet known. As before, thermodynamical
self-consistency will determine its behavior.

The energy density and the pressure are now the sum of two terms referring
to the photon and to the ground state respectively:%
\begin{equation}
\rho _{M}=\rho _{M,\gamma }+\rho _{M,gs},\text{ }p_{M}=p_{M,\gamma
}+p_{M,gs},
\end{equation}%
with:%
\begin{eqnarray}
\rho _{M,\gamma } &=&3\int_{0}^{\infty }\frac{k^{2}dk}{2\pi ^{2}}\frac{\sqrt{%
m_{\gamma }^{2}+k^{2}}}{\exp (\frac{\sqrt{m_{\gamma }^{2}+k^{2}}}{T})-1},%
\text{ }\rho _{M,gs}=4\pi \Lambda _{M}^{3}T \\
p_{M,\gamma } &=&-3\,T\int_{0}^{\infty }\frac{k^{2}dk}{2\pi ^{2}}\ln \left(
1-e^{-\frac{\sqrt{m_{\gamma }^{2}+k^{2}}}{T}}\right) ,\text{ }p_{M,gs}=-\rho
_{M,gs}\,.
\end{eqnarray}%
The corresponding dimensionless quantities are expressed in terms of $%
\lambda =2\pi T/\Lambda _{E}$ and read:%
\begin{eqnarray}
\overline{\rho }_{M,\gamma } &=&\frac{3}{2\pi ^{2}}\int_{0}^{\infty }dx\frac{%
x^{2}\sqrt{x^{2}+b^{2}}}{e^{\sqrt{x^{2}+b^{2}}}-1},\text{ }\overline{\rho }%
_{M,gs}=\left( \frac{\Lambda _{M}}{\Lambda _{E}}\right) ^{3}\frac{2(2\pi
)^{4}}{\lambda ^{3}}  \label{rhoadmag} \\
\overline{p}_{M,\gamma } &=&-\frac{3}{2\pi ^{2}}\int_{0}^{\infty }x^{2}dx\ln
\left( 1-e^{-\sqrt{x^{2}+b^{2}}}\right) ,\text{ }\overline{p}_{M,gs}=-\text{ 
}\overline{\rho }_{M,gs}\,.  \label{padmag}
\end{eqnarray}%
The ratio $\Lambda _{M}/\Lambda _{E}$ is determined by imposing that the
pressure and the photon mass are continuous functions across the phase
transition at $\lambda _{c,E}=$ $13.87$ (second order phase
transition, see \cite{Hofmann2005}). A continuous photon mass requires $b(\lambda
\rightarrow \lambda _{c,E}^{-})=0.$ Thus, by equating eq.\,(\ref{pad}) and
eq.\,(\ref{padmag}), we obtain the following matching condition:%
\begin{equation}
\left( \frac{\Lambda _{M}}{\Lambda _{E}}\right) ^{3}=1+\frac{\lambda
_{c,E}^{3}}{720(2\pi )^{2}}=1.09\,.
\end{equation}%
As a consequence, the energy-gap at the phase transition reads:%
\begin{equation}
\Delta =\overline{\rho }_{M}(\lambda _{c,E})-\overline{\rho }_{E}(\lambda
_{c,E})=\frac{4}{3}\frac{\pi ^{2}}{30}\,.  \label{gapdelta}
\end{equation}

We require thermodynamical self-consistency in the magnetic phase: By
substituting the expressions (\ref{rhoadmag}), (\ref{padmag}) into eq.\,(\ref%
{tsc}) we obtain the following differential equation for $b=b(\lambda )$:%
\begin{equation}
1=-\frac{3\lambda ^{3}}{(2\pi )^{6}}\left( \frac{\Lambda _{E}}{\Lambda _{M}}%
\right) ^{3}\left( \lambda \frac{db}{d\lambda }+b\right) bD(b),\text{ }%
b(\lambda =\lambda _{c,E})=0\,.
\end{equation}%
The behavior of the magnetic coupling $g(\lambda )=b(\lambda )\lambda
^{3/2}/4\pi $ is shown in Fig.\,\ref{Fig1B}. A pole at the critical value $T_{c,M}=11.57\Lambda_E/2\pi$ is encountered. At this temperature 
the photon mass diverges. For temperature smaller than $T_{c,M}$ the system is in a completely confining phase.\vspace{0.1cm}\\ 
{\bf Supercooled electric phase:}\\ 
(Occurs for $\Lambda_{E}\lambda _{\ast c,M}/2\pi=T_{\ast}\leq T\leq
T_{c,E}=\Lambda _{E}\lambda _{c,E}/2\pi,$ $\lambda _{\ast}=12.15$)

The system cooling down from the deconfining into the preconfining phase stays
in a supercooled deconfining phase by virtue of the energy gap. Energy density and pressure in the
supercooled deconfining phase are given as:%
\begin{equation}
\overline{\rho }_{E}^{s.c.}=\overline{\rho }_{E,\gamma }^{s.c.}+\overline{%
\rho }_{E,gs},\text{ }\overline{p}_{E}^{s.c.}=\overline{p}_{E,\gamma
}^{s.c.}+\overline{p}_{E,gs}^{s.c.}\ \ \ \ \ \mbox{where}
\end{equation}%
\begin{equation}
\overline{\rho }_{E,\gamma }^{s.c.}=2\frac{\pi ^{2}}{30}\,,\ \ \text{ }\overline{%
\rho }_{E,gs}=\frac{2(2\pi )^{4}}{\lambda ^{3}}\,;\ \ \ \ \text{ }\overline{p}%
_{E,\gamma }^{s.c.}=2\frac{\pi ^{2}}{90}\,,\ \ \text{ }\overline{p}_{E,gs}^{s.c.}=-%
\text{ }\overline{\rho }_{E,gs}\,.
\end{equation}%
The behavior of $\overline{\rho }_{E}^{s.c.}$ and $\overline{\rho }_{M}$ is
indicated in Fig.\,\ref{Fig-3} where the corresponding transition value $\lambda _{\ast }
$ ($\overline{\rho }_{E}^{s.c.}(\lambda _{\ast })=\overline{\rho }%
_{M}(\lambda _{\ast })$) is determined as $\lambda _{\ast }=12.15$.

\end{document}